\documentclass{sf2a-conf2014}
\usepackage{graphicx}
\usepackage{hyperref}
\usepackage[]{natbib}  
\usepackage{epstopdf}
\usepackage{bm}
\usepackage{amsmath}
\usepackage{amssymb}

\def\BibTeX{{\rm B\kern-.05em{\sc i\kern-.025em b}\kern-.08em
    T\kern-.1667em\lower.7ex\hbox{E}\kern-.125emX}}
\bibpunct{(}{)}{;}{a}{}{,}  

\begin{document}

\TitreGlobal{SF2A 2014}


\title{Inertial waves in differentially rotating low-mass stars\\and tides}

\runningtitle{Inertial waves in differentially rotating low-mass stars and tides}

\author{M. Guenel}\address{Laboratoire AIM Paris-Saclay, CEA/DSM/IRFU/SAp - Universit\'e Paris Diderot - CNRS, 91191 Gif-sur-Yvette, France}

\author{C. Baruteau}\address{IRAP, Observatoire Midi-Pyr\'en\'ees, 14 avenue Edouard Belin, 31400 Toulouse, France}

\author{S. Mathis$^1$}
\author{M. Rieutord$^2$}



\setcounter{page}{237}


\maketitle


\begin{abstract}

Star-planet tidal interactions may result in the excitation of inertial waves in the convective region of stars. Their dissipation plays a prominent role in the long-term orbital evolution of short-period planets. If the star is assumed to be rotating as a solid-body, the waves' Doppler-shifted frequency is restricted to $[-2 \Omega, 2 \Omega]$ ($\Omega$ being the angular velocity of the star) and they can propagate in the entire convective region. However, turbulent convection can sustain differential rotation with an equatorial acceleration (as in the Sun) or deceleration that may modify waves propagation.
We thus explore the properties of inertial modes of oscillation in a conically differentially rotating background flow whose angular velocity depends on the latitudinal coordinate only, close to what is expected in the external convective envelope of low-mass stars. We find that their frequency range is broadened by differential rotation, and that they may propagate only in a restricted part of the envelope. In some cases, inertial waves form shear layers around short-period attractor cycles. In others, they exhibit a remarkable behavior when a turning surface or a corotation layer exists in the star. We discuss how all these cases can impact tidal dissipation in stars.
\end{abstract}



\section{Introduction}

The tidal force exerted by a planet or a stellar companion on its host star may excite inertial waves in the external convective envelope of low-mass stars \citep[see][]{OgilvieLin2007}, which are low-frequency waves whose restoring force is the Coriolis acceleration. The dissipation of the energy and angular momentum carried away by these waves may play an important role on the orbital architecture of planetary systems and on the rotation of their components \citep[see][]{Albrecht2012,Ogilvie2014}, yet it has only recently started being investigated. It has been shown by \cite{Baruteau2013} that differential rotation may strongly affect the propagation and dissipation of linear inertial modes of oscillations. Their study was restricted to shellular and cylindrical rotation profiles, but turbulent convection can also establish conical --- solar/antisolar-type --- differential rotation profiles \citep[see][]{Matt2011,Gastine2014}, as is observed in the Sun. In this work, we explore the propagation and dissipation of linear inertial modes of oscillation in stellar convective envelopes (for low-mass stars only) with conical differential rotation that depends mainly/only on the colatitude.
  
\section{Tidal inertial waves in a differentially rotating shell}

  \begin{figure}[ht!]
  \centering
  \includegraphics[width=0.45\textwidth]{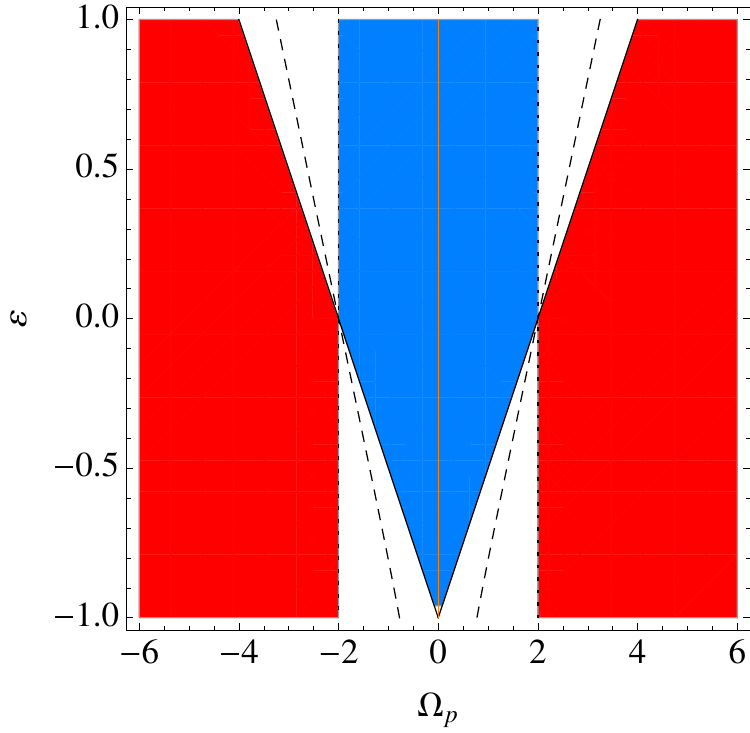}
  \includegraphics[width=0.45\textwidth]{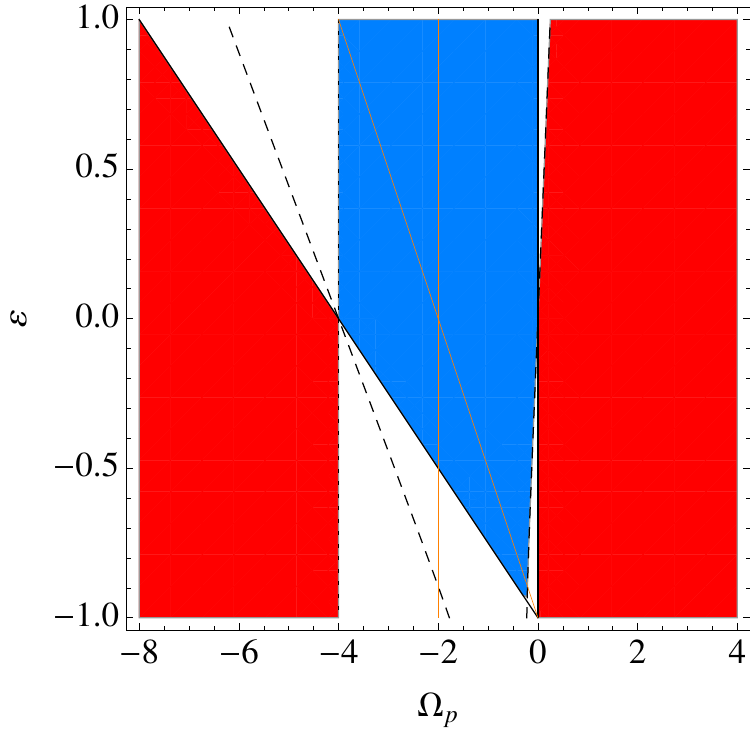}
  \caption{Illustration of the two kinds of inertial modes propagating in a fluid with conical rotation profile $\Omega(\theta)/\Omega_{\rm ref} = 1+\varepsilon \sin^2\theta$. {\bf Left~:} $m=0$. {\bf Right~:} $m=2$. {\bf Blue~:} D modes that propagate in the whole shell ($\xi>0$ everywhere). {\bf White~:} DT modes that exhibit at least one turning surface inside the shell (the sign of $\xi$ changes in the shell). {\bf Red~:} No inertial modes may propagate ($\xi<0$ everywhere). {\bf Orange~:} The modes in the region between the two orange lines feature a corotation layer inside the shell.} 
  \label{fig:BBR}
  \end{figure}

\subsection{The set-up}
We model the convective envelope of a low-mass star as a rotating homogeneous incompressible viscous fluid inside a spherical shell of aspect ratio $\eta$ and external radius $R$. The fluid is assumed to have a conical differential rotation profile \emph{i.e.} depending only on the colatitudinal coordinate $\theta$ :
\begin{equation}
\Omega(\theta) = \Omega_{\rm ref}\left(1 + \varepsilon \sin^2\theta\right) = \Omega_{\rm ref}\left(1 + \varepsilon \frac{s^2}{s^2+z^2}\right), \quad \mbox{with} -R \leqslant s,z \leqslant R. \\
\end{equation}
We use standard cylindrical coordinates $(s,\phi,z)$, with the z-axis aligned with the rotation axis. $\Omega_{\rm ref}$ denotes the angular velocity at the rotation axis while $\varepsilon$ gives the behavior of the differential rotation, \emph{i.e.}:
\begin{itemize}
\item $\varepsilon > 0$ that corresponds to the case of solar differential rotation (equatorial acceleration),
\item $\varepsilon < 0$ for anti-solar differential rotation (polar acceleration).
\end{itemize}

For the sake of simplicity, we ignore the non-linear interactions between the different waves, as well as the interaction between waves and convection. In an inertial frame, we look for velocity ($\mathbf u$) and reduced pressure ($p$) perturbations proportional to $\exp\left(i\Omega_p t + im\phi\right)$ that satisfy the linear system :
\begin{equation}
\begin{cases}
& i \tilde{\Omega}_p {\bf u} + 2 \Omega {\mathbf e_{\bm z}} \times {\mathbf u} + s \left( {\mathbf u}\cdot\nabla\Omega \right) {\mathbf e_{\bm \phi}} = -{\nabla}p + \nu \nabla^2 {\mathbf u}\\
&\nabla \cdot {\mathbf u} = 0
\end{cases},
\label{eq:momentum}
\end{equation}
where $\tilde{\Omega}_p(\theta) = \Omega_p+m\Omega(\theta)$ is the Doppler-shifted frequency of the mode, $\nu$ is the viscosity and $(s,z,\phi)$ are the cylindrical coordinates defined above. We may use both rigid and/or stress-free boundary conditions at the inner and outer boundaries of the shell.

\subsection{Modification of inertial waves by conical differential rotation}
  
Combining the linearized hydrodynamical equations and using the short-wavelength approximation, we get the following Poincar\'e equation for the pressure perturbation $p$ only:
\begin{equation}
\frac{\partial^2 p}{\partial s^2} + \frac{A_z}{\tilde{\Omega}_p^2}\frac{\partial^2 p}{\partial s \partial z}+\left( 1-\frac{A_s}{\tilde{\Omega}_p^2}\right) \frac{\partial^2 p}{\partial z^2}=0,
\label{eq:poincare}
\end{equation}
where
\begin{equation}
A_s(\theta) = \frac{2\Omega}{s}\frac{\partial}{\partial s}(s^2 \Omega) \quad \mbox{and}\quad A_z(\theta) = \frac{2\Omega}{s}\frac{\partial}{\partial z}(s^2 \Omega).
\end{equation}

\begin{figure}[ht!]
\centering
\includegraphics[width=0.45\textwidth]{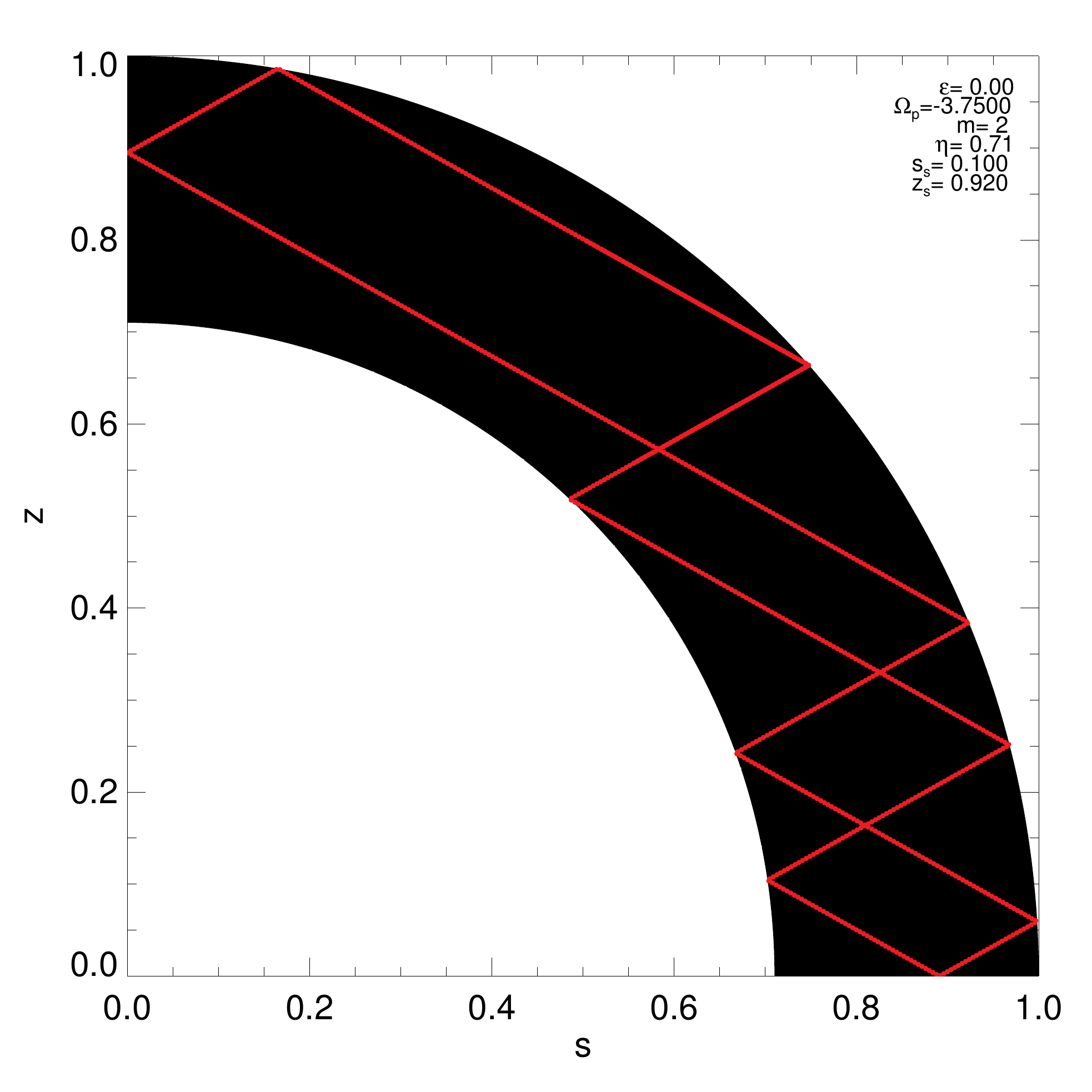}
\includegraphics[width=0.45\textwidth]{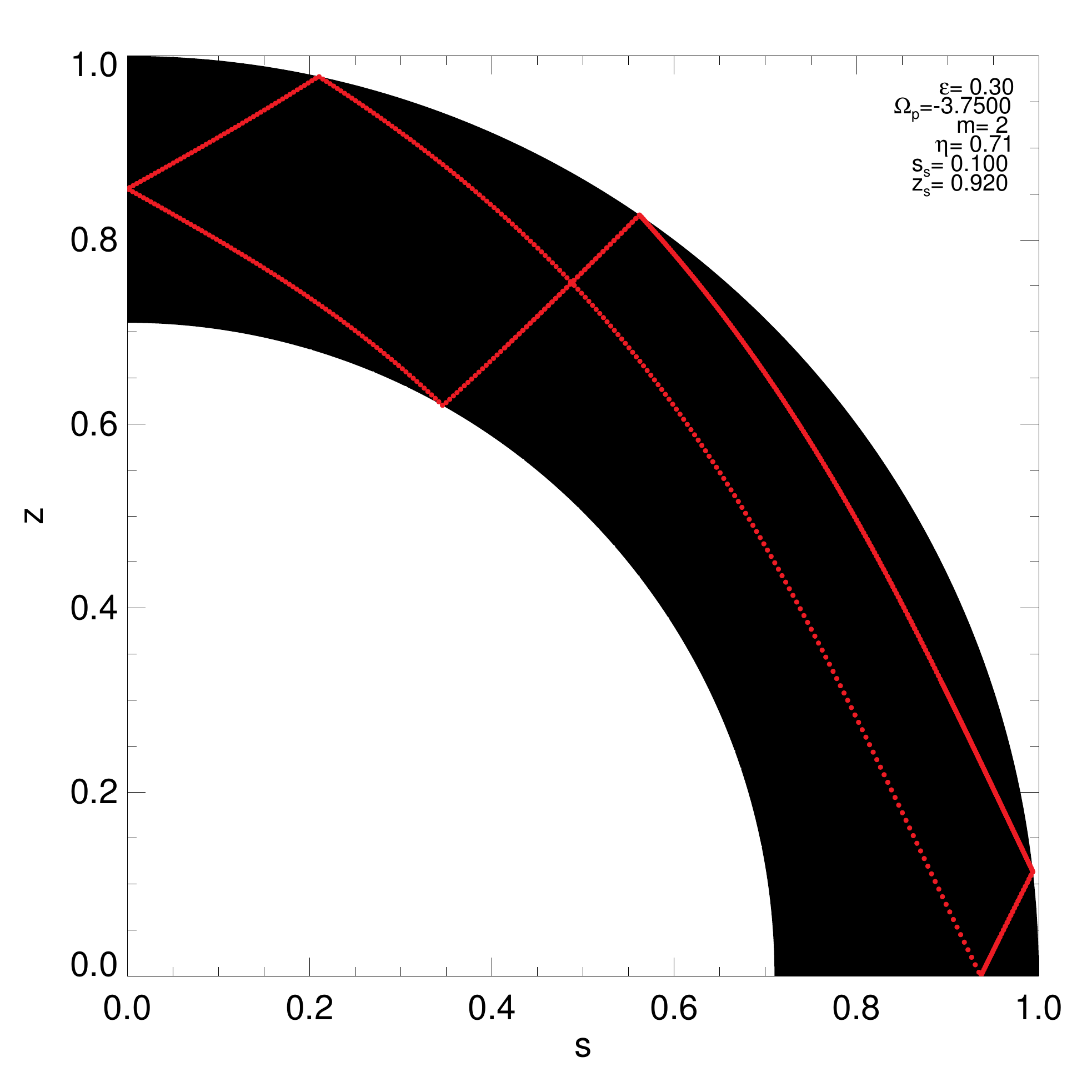}
\includegraphics[width=0.45\textwidth]{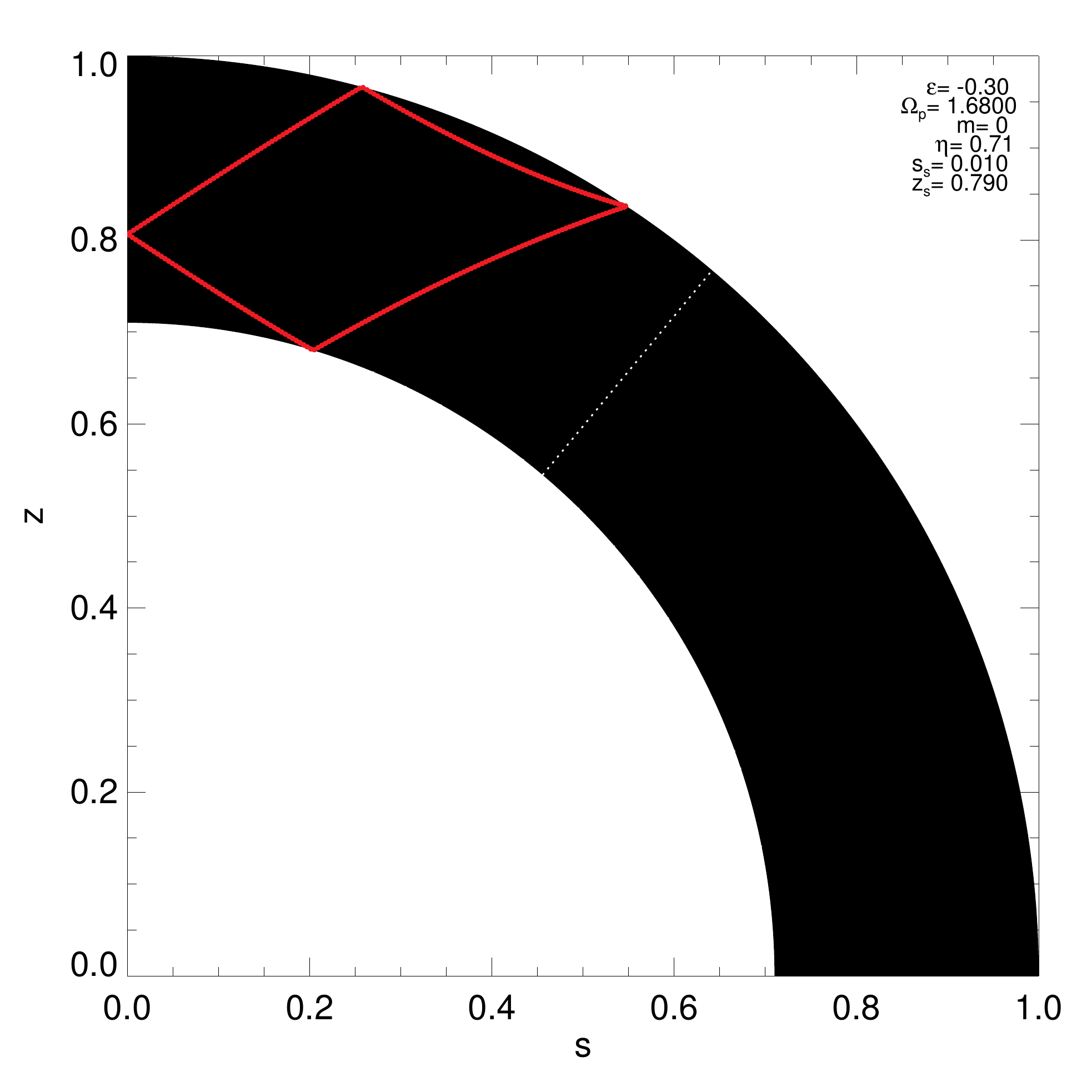}
\includegraphics[width=0.45\textwidth]{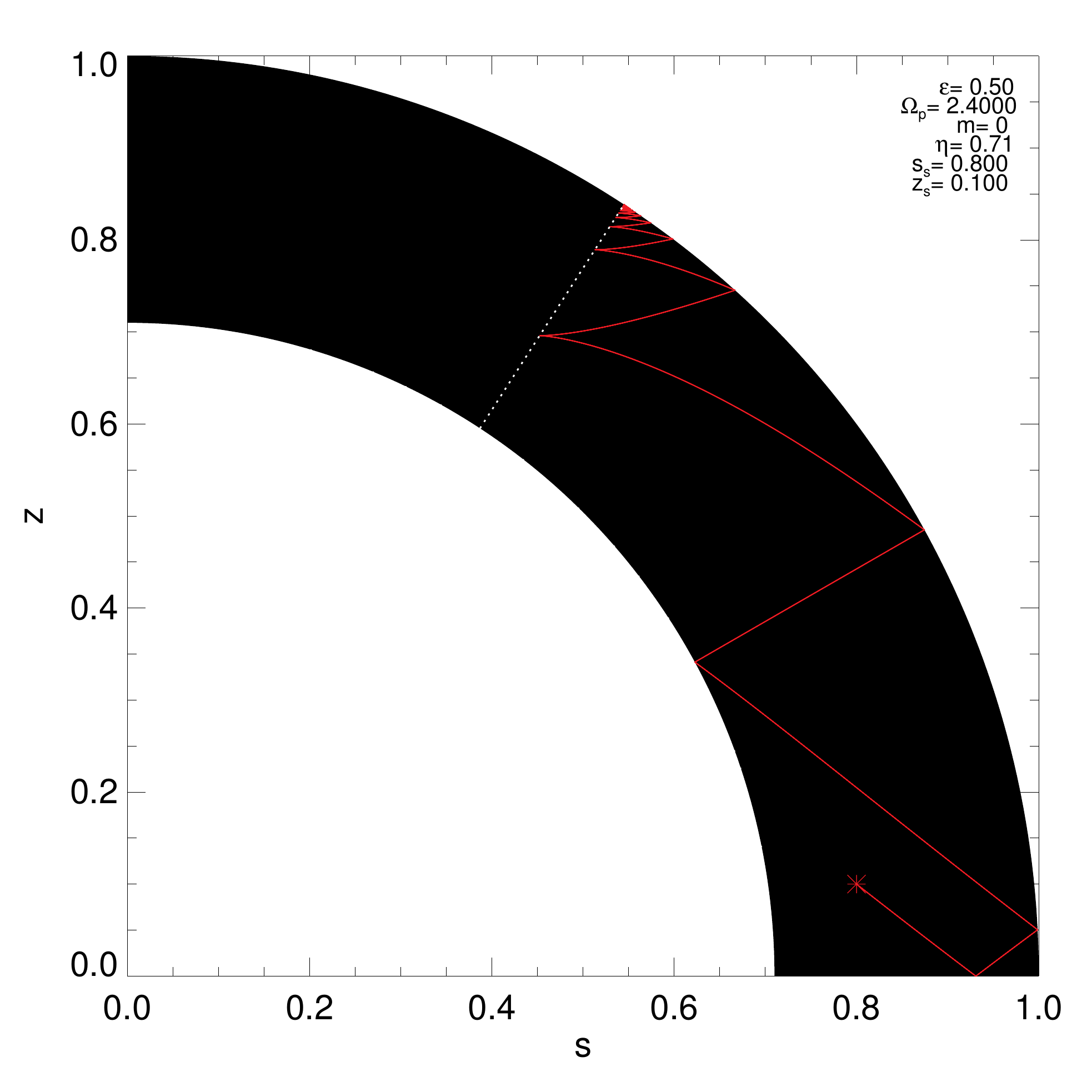}
\caption{{\bf Upper left~:} Example of an attractor cycle of the path of characteristics for a D mode with $m=2$, $\Omega_p/\Omega_{\rm ref}=-3.75$, solid-body rotation ($\varepsilon=0$) and the solar aspect ratio $\eta=0.71$. {\bf Upper right~:} Same for a D mode in a solar-like convective envelope ($\varepsilon=0.3$). {\bf Lower left~:} Same for a DT mode with $m=0$, frequency $\Omega_p/\Omega_{\rm ref}=1.68$ and anti-solar conical rotation ($\varepsilon = -0.3$), the dotted line representing the turning surface. {\bf Lower right~:} Illustration of the focusing of the paths of characteristics at the intersection of the turning surface (dotted line) and the outer boundary of the shell ($m=0$, $\Omega_p/\Omega_{\rm ref}=2.4$ and $\varepsilon=0.5$).}
\label{fig:attractors}
\end{figure}

Rayleigh's stability criterion requires $A_s>0$ everywhere in the shell, which yields $\varepsilon \geqslant -1$. Note also that in the case of solid-body rotation, we obtain $A_s = 4\Omega^2$ and $A_z = 0$. This equation is hyperbolic and waves propagate when the discriminant $\displaystyle\xi(\theta) = A_z^2 + 4 \tilde{\Omega}_p^2 \left( A_s - \tilde{\Omega}_p^2 \right)$ is positive. In that case the equation governing the path of characteristics in a meridional plane, along which energy propagates, reads :
\begin{equation}
\frac{dz}{ds} = \frac{1}{2\tilde{\Omega}_p^2} \left( A_z \pm \xi^{1/2}\right).
\end{equation}
If $\xi$ vanishes somewhere in the shell, the paths of characteristics bounce off ``turning surfaces" where $\xi=0$ (see the lower plots of fig. \ref{fig:attractors}). Moreover, corotation resonances occur when $\tilde{\Omega}_p = 0$.
  
The study of the Poincar\'e equation \ref{eq:poincare} allows to identify waves modification by the conical differential rotation. First, the usual frequency-range $\tilde{\Omega}_p \in \left[-2\Omega_{\rm ref},2\Omega_{\rm ref}\right]$ can be broadened by differential rotation as shown in fig. \ref{fig:BBR}.
When $\varepsilon\neq 0$ we find that the slope of characteristics now depends on $(s,z)$, which means their paths are not straight lines anymore but curved lines (see the upper plots of fig. \ref{fig:attractors}),. Additionally, some modes can be trapped (latitude-wise) because of the existence of a turning surface at which $\xi$ vanishes (see lower left-hand plot of fig. \ref{fig:attractors}). Following the same terminology as in \cite{Baruteau2013}, we call them DT modes (D for differential rotation, and T for turning surface) in opposition to the D modes which propagate in the entire shell.
When $m\neq0$, corotation layers may exist in the shell where $\tilde{\Omega}_p = 0$, with locally vertical paths of characteristics.
Paths of characteristics may still converge towards short-period cycles or ``attractors", as in the case of solid-body rotation, but we find that they may also focus at the intersection of a turning surface with the boundaries of the shell as illustrated by the lower right-hand plot of fig. \ref{fig:attractors}.

\subsection{First numerical solutions}

\begin{figure}[ht!]
\centering
\includegraphics[width=0.45\textwidth]{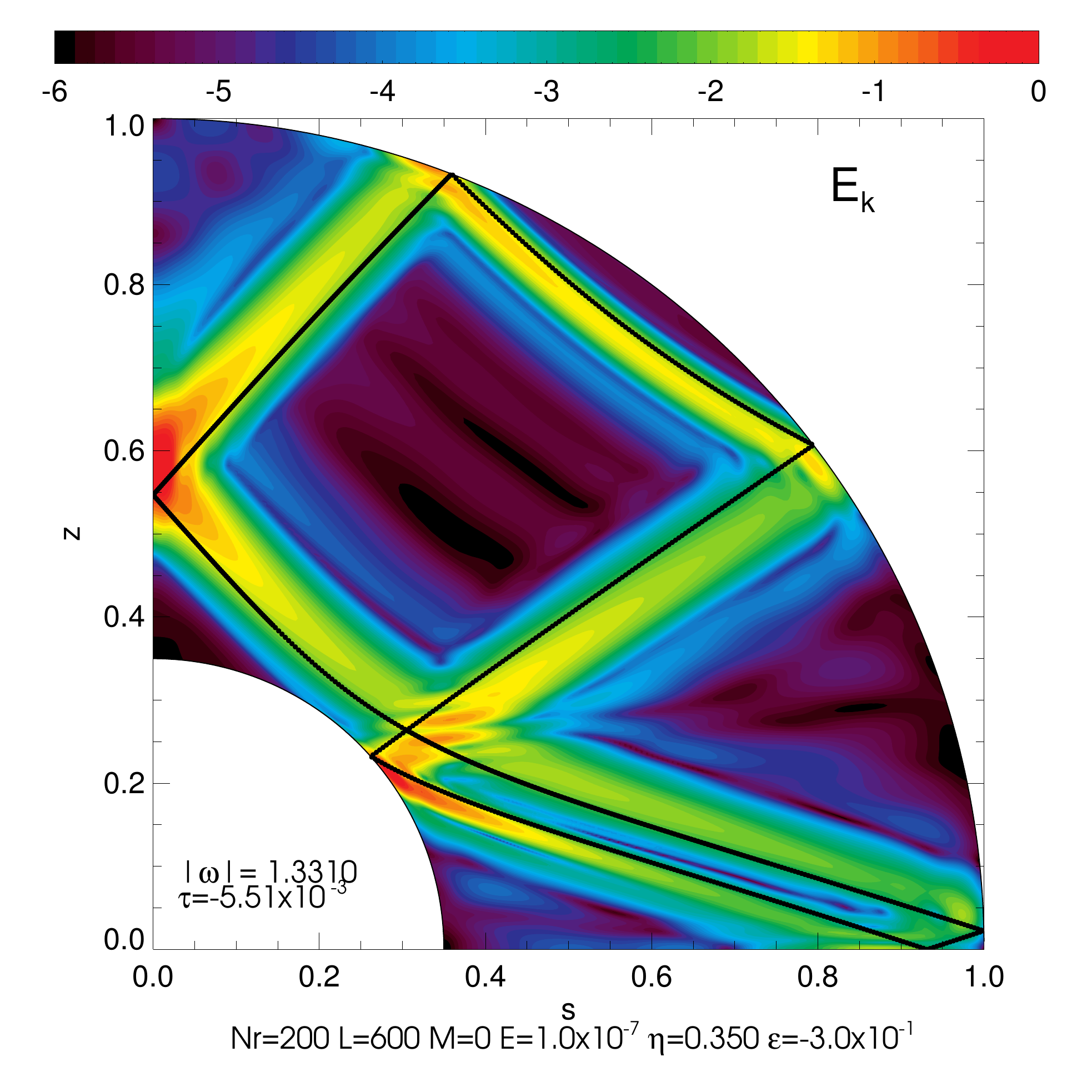}
\includegraphics[width=0.45\textwidth]{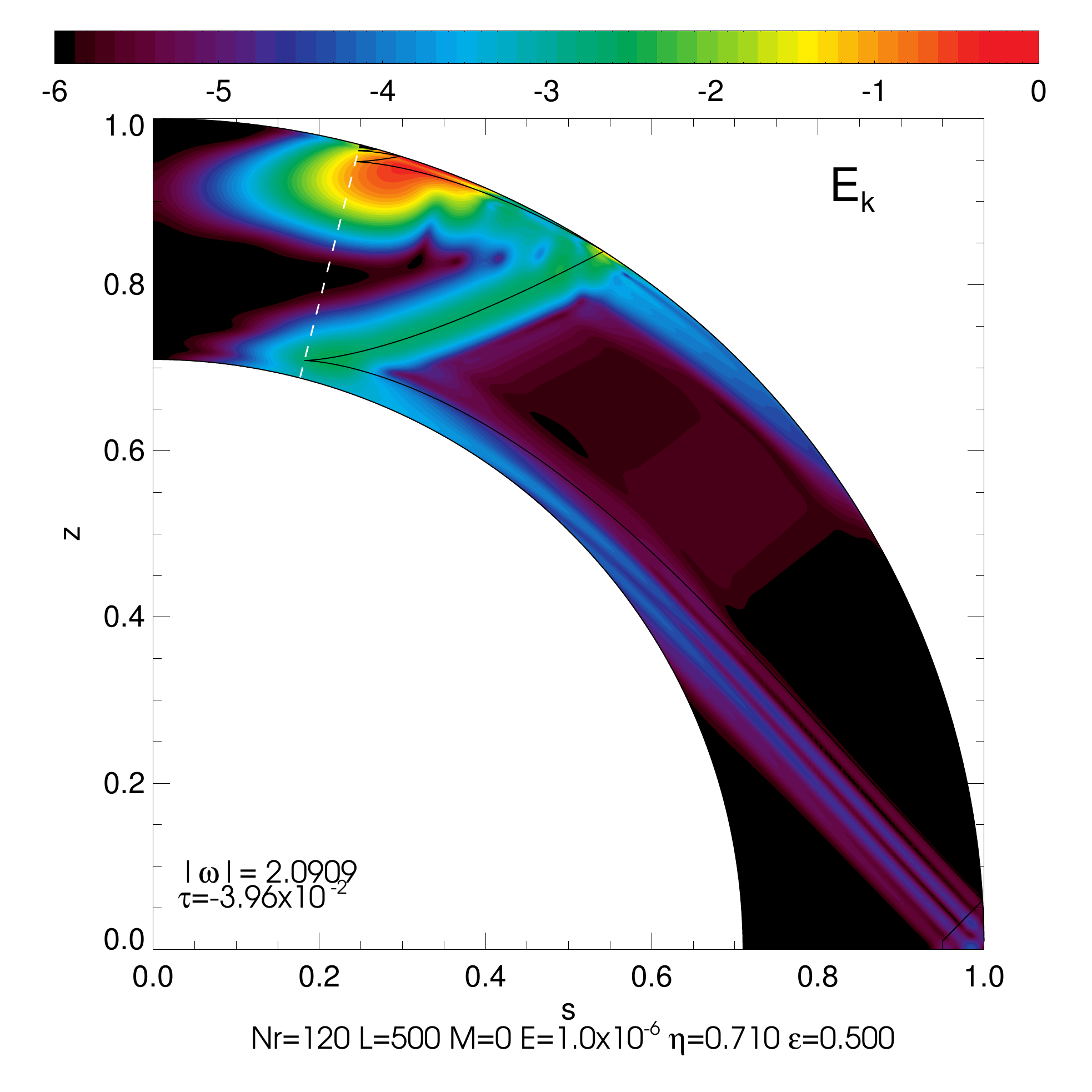}
\caption{{\bf Left~:} Kinetic energy of a D mode with anti-solar differential rotation ($m=0$, $\Omega_p/\Omega_{\rm ref}=1.33$, $\eta=0.35$, $\varepsilon=-0.3$ and $E=10^{-7}$) . Note that in the Sun, $\varepsilon \approx 0.3$. {\bf Right~:} Kinetic energy of a DT mode in a solar-like convective envelope ($m=0$, $\Omega_p/\Omega_{\rm ref}=2.09$, $\eta=0.71$, $\varepsilon=0.5$ and $E=10^{-6}$). The focusing of the path of characteristics around $(s,z)=(0.25,0.95)$ creates an accumulation of kinetic energy around this region.}
\label{fig:modes}
\end{figure}

We solve Eqs. (\ref{eq:momentum}) along with stress-free boundary conditions using a spectral code \citep[see details in][]{Rieutord1987}. We show in fig. \ref{fig:modes} two examples of linear modes of oscillations in the case of conical differential rotation for $m=0$ for different Ekman numbers, defined by $E=\nu/(\Omega_{\rm ref}R^2)$ in our framework. Notice that the agreement between the structure of the kinetic energy of the mode (in colors) and the paths of characteristics (overplotted in black) is good, which confirms the validity of our results. In a near future, we will carry out the same type of calculations for tidally-forced inertial modes, which will allow us to evaluate the tidal dissipation which is due to inertial waves in the convective envelope of differentially rotating low-mass stars.

\section{Conclusions}

This work is a first step towards the understanding of the physics of tidally excited inertial waves under conditions that are close to those of differentially rotating external convective envelopes of low-mass stars. The next step will be to introduce a forcing term in the equation of momentum in order to simulate the action of a companion.

\bibliographystyle{aa}  
\bibliography{guenel1} 

\end{document}